# Interlayer magnetophononic coupling in MnBi$_2$Te$_4$

Hari Padmanabhan,[α,*,1] Maxwell Poore,[*,2] Peter Kim,[*,2] Nathan Z. Koocher,[3] Vladimir A. Stoica,[1] Danilo Puggioni,[3] Huaiyu Wang,[1] Xiaozhe Shen,[4] Alexander H. Reid,[4] Mingqiang Gu,[3] Maxwell Wetherington,[1] Seng Huat Lee,[5,6] Richard Schaller,[7] Zhiqiang Mao,[5,6] Aaron M. Lindenberg,[8] Xijie Wang,[4] James M. Rondinelli,[3] Richard Averitt,[β,2] Venkatraman Gopalan[γ,1]

[1]Materials Research Institute and Department of Materials Science & Engineering, Pennsylvania State University, University Park, Pennsylvania 16802, USA
[2]Department of Physics, University of California San Diego, La Jolla, California 92093, USA
[3]Department of Materials Science and Engineering, Northwestern University, Evanston, Illinois 60208, USA
[4]SLAC National Accelerator Laboratory, Menlo Park, California 94025, USA
[5]2D Crystal Consortium, Materials Research Institute, Pennsylvania State University, University Park, Pennsylvania 16802, USA
[6]Department of Physics, Pennsylvania State University, University Park, Pennsylvania 16802, USA
[7]Center for Nanoscale Materials, Argonne National Laboratory, Lemont, Illinois 60439, USA
[8]Department of Materials Science and Engineering, Stanford University, Stanford, CA, USA

*These authors contributed equally to this work.
[α]hari@psu.edu
[β]raveritt@ucsd.edu
[γ]vgopalan@psu.edu

**The emergence of magnetism in quantum materials creates a platform to realize spin-based applications in spintronics, magnetic memory, and quantum information science. A key to unlocking new functionalities in these materials is the discovery of tunable coupling between spins and other microscopic degrees of freedom. We present evidence for interlayer magnetophononic coupling in the layered magnetic topological insulator MnBi$_2$Te$_4$. Employing magneto-Raman spectroscopy, we observe anomalies in phonon scattering intensities across magnetic field-driven phase transitions, despite the absence of discernible static structural changes. This behavior is a consequence of a magnetophononic wave-mixing process that allows for the excitation of zone-boundary phonons that are otherwise 'forbidden' by momentum conservation. Our microscopic model based on density functional theory calculations reveals that this phenomenon can be attributed to phonons modulating the interlayer exchange coupling. Moreover, signatures of magnetophononic coupling are also observed in the time domain through the ultrafast excitation and detection of coherent phonons across magnetic transitions. In light of the intimate connection between magnetism and topology in**

**MnBi$_2$Te$_4$, the magnetophononic coupling represents an important step towards coherent on-demand manipulation of magnetic topological phases.**

The realization of magnetic order in functional quantum materials creates a rich platform for the exploration of fundamental spin-based phenomena, as exemplified in strongly correlated materials[1], multiferroics[2], and more recently, magnetic topological materials[3]. As such, these materials hold great promise for application in spintronics, magnetic memory, and quantum information technology. A new paradigm has recently emerged with the discovery of atomically thin magnets, derived from layered, quasi-two-dimensional materials[4]. In such materials, magnetic order is characterized by strongly anisotropic exchange interactions, with interlayer exchange coupling that is an order-of-magnitude weaker than the in-plane exchange coupling. The weak interlayer exchange coupling offers a high degree of tunability in the two-dimensional limit, enabling the realization of phenomena such as magnetic switching via electric fields[5] and electrostatic doping[6]. Such tunability could potentially be made even more potent in combination with additional functionalities such as those outlined above. For instance, the Mn(Bi,Sb)$_{2n}$Te$_{3n+1}$ family of layered antiferromagnets is the first experimental realization of intrinsic magnetic order in topological insulators[7–9]. The interlayer magnetic order is intimately connected to the band topology, with experimental demonstration of switching between quantum anomalous Hall and axion insulator states[10], and realization of a field-driven Weyl semimetal state[11]. In this context, the discovery of new, efficient coupling pathways between the interlayer exchange and other microscopic degrees of freedom would not only add to the rich spectrum of low-dimensional magnetic phenomena, but also potentially unlock pathways for the dynamic manipulation of magnetism and band topology.

In this work, we observe that interlayer magnetic order in MnBi$_2$Te$_4$ is strongly coupled to phonons, manifesting in the optical excitation of zone-boundary phonons that are otherwise forbidden due to the conservation of momentum. This novel magnetophononic response is a consequence of a coherent wave-mixing process between the antiferromagnetic order and $A_{1g}$ optical phonons, as determined from equilibrium and time-domain spectroscopy across temperature- and magnetic field-driven phase transitions.

Our microscopic model based on first-principles calculations reveals that this phenomenon can be attributed to phonons modulating the interlayer exchange coupling.

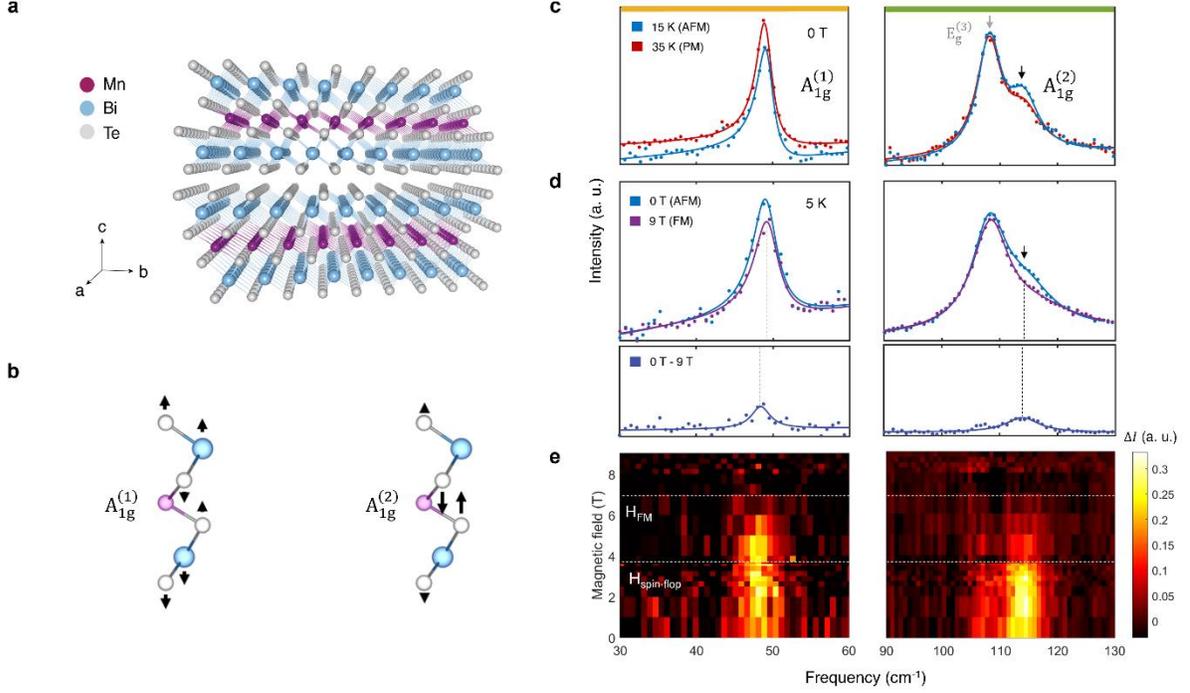

**Fig. 1. Phonon anomalies across magnetic phase transitions in MnBi$_2$Te$_4$. a,** Crystal structure of MnBi$_2$Te$_4$. **b,** Eigendisplacements of the $A_{1g}^{(1)}$ and $A_{1g}^{(2)}$ modes, with arrows denoting displacement of ions. **c,** Raman spectra of $A_{1g}^{(1)}$ (left) and $A_{1g}^{(2)}$ (right) modes in the paramagnetic (PM) and antiferromagnetic (AFM) phases at 0 T, shown in red and blue respectively. **d,** Raman spectra of $A_{1g}^{(1)}$ (left) and $A_{1g}^{(2)}$ (right) modes in the AFM and ferromagnetic (FM) phases at 5 K, shown in blue, and purple respectively. The lower panels show the difference between spectra in the AFM and FM phases. **e,** Contour plots of the difference upon subtracting the 9 T spectrum, as a function of magnetic field. The dotted lines denote the FM and spin-flop critical fields.

MnBi$_2$Te$_4$ exhibits magnetic order below a temperature of $T_N = 24$ K, with in-plane ferromagnetic coupling, and out-of-plane antiferromagnetic (AFM) coupling[12], as shown in Fig. 1a. With an applied out-of-plane magnetic field, a spin-flop transition occurs at 3.7 T, developing into a fully polarized ferromagnetic-like state (FM) at a critical field of 7.7 T[12]. We first present measurements of the phonon spectra across the magnetic phase transitions in MnBi$_2$Te$_4$, using magneto-Raman spectroscopy. The full polarized Raman phonon spectra, selection rules, and peak assignments can be found in Supplementary Information Section S1. Our peak assignment is fully consistent with a previous study[13] that investigated Raman phonons in thin flakes of MnBi$_2$Te$_4$ as a function of number of layers. Here we focus on two fully symmetric '$A_{1g}$' phonon modes at frequencies of 49 cm$^{-1}$ and 113 cm$^{-1}$, labeled $A_{1g}^{(1)}$ and $A_{1g}^{(2)}$ respectively The phonon eigendisplacements, calculated using density functional theory (DFT) simulations, are shown in Fig. 1b. Representative spectra at 0 T, in the AFM phase at 15 K and the paramagnetic (PM) phase at 35 K, are shown in Fig. 1c. We observe that the $A_{1g}^{(2)}$ mode clearly exhibits an anomalous increase in scattering intensity in the AFM phase, which has not been reported in previous studies[13]. The temperature-dependence of the $A_{1g}^{(1)}$ mode is discussed in detail in SI Section 2. In the following, we focus on the magnetic field-dependent behavior. At a magnetic field of 9 T, where MnBi$_2$Te$_4$ is in the fully polarized ferromagnetic (FM) state, the spectral weight of both modes decreases, as shown in the top panel of Fig. 1d. This is highlighted by subtracting the spectrum at 9 T from the spectrum at 0 T and plotting the residual in the bottom panel of Fig. 1d. In Fig. 1e, the residual is plotted as a function of magnetic field $H$, upon subtracting the 9 T spectrum. A clear correlation is observed between the residual scattering intensity of the $A_{1g}$ modes and the critical magnetic fields for the spin-flop and FM transitions, denoted by dashed white lines.

The fractional change in integrated intensity of the $A_{1g}^{(2)}$ mode is plotted as a function of temperature in Fig. 2a (green dots). The integrated intensity follows the AFM order parameter, tracked by the (1 0 5/2) neutron diffraction Bragg peak[14] (purple dots). The grey line is a fit to the power law $I \propto \left(1 - \frac{T}{T_N}\right)^{2\beta}$, with $\beta = 0.35$ as in the reference[14]. Furthermore, a plot of the scattering intensity of the $A_{1g}^{(1)}$ and $A_{1g}^{(2)}$ modes (Fig. 2b) as a function of magnetic field reveals the fractional change in integrated intensities of both modes tracks the AFM order parameter[15] across the spin-flop transition at 3.7 T, and into the fully polarized ferromagnetic state above 7.7 T. The integrated intensities of the $A_{1g}^{(1)}$ and $A_{1g}^{(2)}$ modes increase by fractions of 0.15 and 0.3 respectively, in the AFM phase, as compared to the FM phase at 9 T. Additionally, the fractional increase in the $A_{1g}^{(2)}$ intensity as estimated from the PM to AFM transition and FM to AFM transition in Figs. 2a and 2b, respectively, are of the same magnitude, pointing to a common origin.

Importantly, within the limits of our experimental uncertainty (error bars in plots), we do not observe such large changes in the integrated intensity on any of the other Raman phonons (see SI Section S3 for detailed field-dependent data). Below, we show that the experimentally observed temperature- and field-dependent evolution of scattering intensity is consistent with the excitation of 'forbidden' zone-boundary modes of the $A_{1g}^{(1)}$ and $A_{1g}^{(2)}$ phonon branches.

The AFM order along the out-of-plane direction (crystallographic c-axis) results in a magnetic unit cell that is double the size of the crystallographic unit cell, as shown in Fig. 3a. In contrast, in the high-field FM state (and the paramagnetic state), the magnetic unit cell is identical with the crystallographic unit cell, as in the paramagnetic state. This behavior manifests in the anomalous field-dependent scattering intensity of the $A_{1g}$ modes, which follows the AFM order parameter with the magnetic unit cell doubling resulting in a folding of the phonon Brillouin zone, allowing for the optical detection of zone-boundary phonon modes. DFT simulations of the phonon dispersion along the out-of-plane direction reveal a flat dispersion for the $A_{1g}^{(2)}$ mode, and a small dispersion for the $A_{1g}^{(1)}$, consistent with the weak interlayer van der Waals interaction, and our experimental results, denoted in Fig. 3b using bold circles. This supports our assignment of the anomalous scattering intensity as zone-boundary modes. We also consider and rule out alternative explanations for the observed temperature- and magnetic field-dependent scattering intensity changes, such as resonant Raman effects (see Supplementary Information Section S5) and possible magnon resonances overlapping with the considered phonons (see Supplementary Information Section S6).

Magnetic unit cell doubling resulting in the activation of zone boundary phonons is unexpected given the absence of a structural phase transition. Refinement based on neutron diffraction at 10 K and 100 K shows no structural unit cell doubling across the AFM transition, and no changes to the unit cell coordinates to within $10^{-3}$ of the lattice parameters[14]. The negligible change in the spectra of other Raman phonons in $MnBi_2Te_4$ is also consistent with the absence of a structural transition of any kind, and points instead to a mechanism that is mode-dependent.

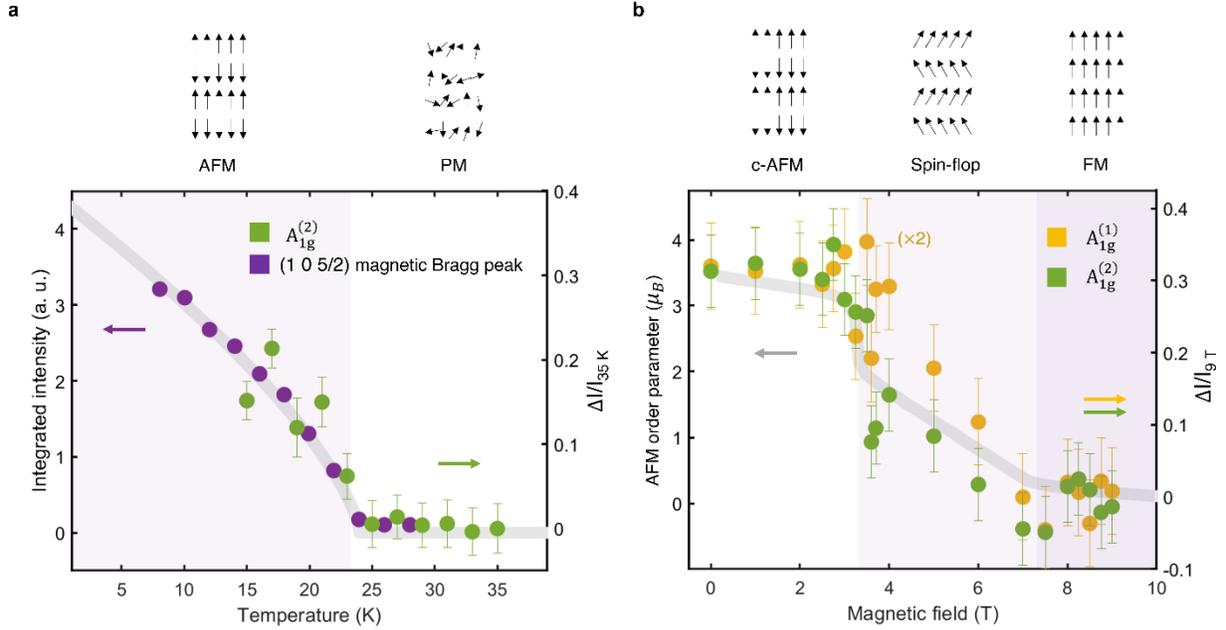

**Fig. 2. Phonon intensities track the antiferromagnetic order parameter. a,** Temperature-dependent fractional change in integrated intensity, $\Delta I/I_{35\,K}$, of the $A_{1g}^{(2)}$ mode, overlayed on integrated intensity of the (1 0 5/2) neutron diffraction peak from reference[14]. The grey line is a fit to $A(1 - T/T_N)^{2\beta}$, with $\beta = 0.35$, $T_N = 24$ K. **b,** The field-dependent fractional change in integrated intensity, $\Delta I/I_{9\,T}$, of the $A_{1g}^{(1)}$ and $A_{1g}^{(2)}$ modes. The grey line is the AFM order parameter, given by $M - 4.5\,\mu_B$, where $M$ is the magnetization measured by magnetometry from reference[15]. Error bars are standard deviations in fit values.

In general, zone-boundary modes are optically inactive or 'forbidden' due to the conservation of crystal momentum. Photons in the visible part of the spectrum have negligible momentum in comparison with the crystal Brillouin zone, and thus momentum conservation dictates that only zero momentum (i. e. zone-center) excitations can be generated and detected in first-order scattering processes. This is shown schematically for Raman scattering in Fig. 3c. This selection rule can be overcome in the presence of other finite momentum waves in the crystal, as observed for instance in the case of structural distortions that double the crystallographic unit cell[16–18]. However, as noted above, MnBi$_2$Te$_4$ does not exhibit any structural transition. Instead, we propose that the crystal momentum is provided by the AFM order, via a magnetophononic wave-mixing process. This is shown schematically in Fig. 3c, where the AFM crystal momentum $q_{AFM} = 2\pi/2c$ interacts with the phonon crystal momentum, allowing for the excitation of zone-boundary ($q = \pi/c$) phonons.

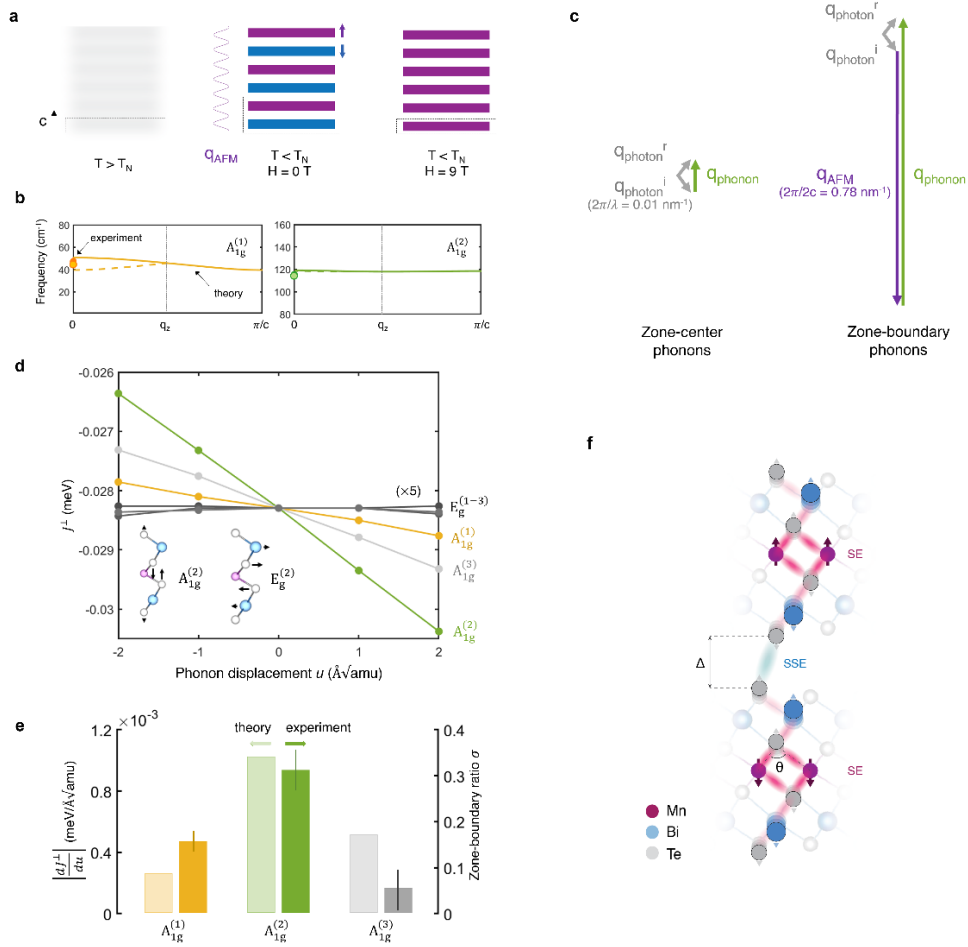

**Fig. 3. Magnetophononic wave-mixing. a,** Schematic of layered magnetic ordering in MnBi$_2$Te$_4$, with blue and purple denoting opposite in-plane spin orientations, and grey denoting disordered spins. The antiferromagnetic (AFM) wavevector is shown schematically, labeled 'q$_{AFM}$'. **b,** The dispersion relations of the $A_{1g}^{(1)}$ and $A_{1g}^{(2)}$ modes along the c-axis, calculated using density functional theory. The experimental zone-center and zone-boundary phonon frequencies are denoted using colored and empty circles respectively. **c,** Schematic of wave-mixing for zone-center and zone-boundary modes. The wavevectors of the photon (i = incident, r = reflected), phonon, and AFM spin-wave are shown using grey, green, and purple arrows (not drawn to scale). **d,** Modulation of the interlayer exchange coupling $J^\perp$ by Raman phonons. Inset shows the eigendisplacements of two representative phonons. **e,** Comparison between the calculated magnetophononic scattering cross-section $\left|\frac{dJ^\perp}{du}\right|$ and the experimental zone-boundary ratio $\sigma$ (see text for definition). Error bars are standard deviations in fit values. **f,** Schematic of superexchange (SE) and super-superexchange (SSE), with Δ denoting the interlayer distance, $\theta$ denoting the Mn-Te-Mn bond angle, and pink and blue clouds denoting SE and SSE pathways, respectively.

Magnetophononic wave-mixing requires a sufficiently strong scattering cross-section to be observable. This scattering cross-section can typically be written in terms of an interaction term in the free energy. For example, the Raman scattering process is due to a coupling of the incident ($E^i$) and reflected ($E^r$) electric fields to a distortion $u$ along a phonon normal mode, via the susceptibility $\chi_e$ (i. e. $F = \left(\frac{d\chi_e}{du}u\right)E^i E^r$). In the case of a finite-momentum structural distortion, phonons couple to the structural distortion through elastic interactions. Analogously, in our model of magnetophononic wave-mixing, phonons couple to the AFM order by modulating the interlayer exchange interaction $J^\perp$. The corresponding interaction term in the free energy can be obtained by first writing down a Heisenberg-like Hamiltonian for the spin energy, $H = \sum_{ij} J_{ij} S_i \cdot S_j$, where $J_{ij}$ is the exchange coupling between spins at sites $i$ and $j$. Since the coupling is to an out-of-plane antiferromagnetic spin wave, we focus on the interlayer (out-of-plane) exchange coupling $J^\perp$ (only nearest-neighbor interlayer interactions are considered). If a phonon modulates the interlayer exchange interaction, the perturbed exchange coupling $J^{\perp\prime}$ can be written as

$$J^{\perp\prime} = J^\perp + \frac{dJ^\perp}{du}u + \cdots \quad (1)$$

Equation 1 is a special case of what is broadly referred to in the literature as 'spin-phonon coupling' (see SI Section S4 for the interpretation of higher-order terms in terms of phonon frequency renormalization). Based on this, the free energy term that couples the antiferromagnetic spin wave to the phonon is, to first order,

$$F = \left(\frac{dJ^\perp}{du}u\right)\sum_i S_i S_{i+1}, \quad (2)$$

where $i$ and $i + 1$ correspond to nearest-neighbor spin pairs in the out-of-plane direction. It is clear that the magnitude of this coupling directly depends on $\frac{dJ^\perp}{du}$. In other words, a magnetophononic wave-mixing is possible only when the phonon mode under consideration sufficiently modulates the interlayer exchange coupling.

A microscopic basis for this model can be obtained using DFT simulations. We simulate the modulation of the interlayer exchange coupling $J^\perp$ by the six Raman phonons of MnBi$_2$Te$_4$, which include three $A_{1g}$ modes (pure out-of-plane eigendisplacements), and three $E_g$ modes (pure in-plane eigendisplacements, see SI Fig. S1b for eigendisplacements). The results, shown in Fig. 3d, indicate a striking dichotomy between the out-of-plane $A_{1g}$ modes and in-plane $E_g$ modes. The $A_{1g}$ modes exhibit an order-of-magnitude larger modulation of $J^\perp$ than the $E_g$ modes. Furthermore, the $A_{1g}^{(2)}$ mode has by far the largest influence on $J^\perp$, consistent with our experimental observation of zone-boundary scattering intensity. A quantitative comparison of this model with our experimental results is possible. This is accomplished by defining an experimental magnetophononic scattering cross-section $\sigma$, as the ratio of the integrated intensity of the zone-boundary mode (i. e. the residual spectra in Fig. 1d) to that of the zone-center mode (spectra at 9 T in Fig. 1d). The scattering cross-section is compared to the calculated interaction term, $\left|\frac{dJ^\perp}{du}\right|$. The plotted results in Fig. 3e show a good agreement between theory and the experiment. In particular, the model reproduces the experimental observation of the $A_{1g}^{(2)}$ mode exhibiting the largest zone-boundary scattering intensity. We note that no signature of a zone-boundary mode was observed in the $A_{1g}^{(3)}$ branch within the experimental uncertainty (see SI Section S3). Finally, also in agreement with the theoretical prediction, no $E_g$ zone-boundary modes were experimentally observed, i. e. $\sigma = 0$ for all $E_g$ modes, within the experimental uncertainty (see SI Section S3).

The theoretical results outlined above can be rationalized in terms of microscopic interlayer exchange pathways. In general, the exchange coupling across a van der Waals (vdW) gap is understood to be the result of a process named 'super-superexchange' (SSE)[19]. In SSE, given that the interlayer exchange interaction is usually much weaker than the intralayer exchange interaction, the two can be effectively decoupled. The individual quasi-two-dimensional layers are treated as macroscopic magnetic moments established by the intralayer superexchange (shown in pink in Fig. 3f), which couple across the vdW gap via the weaker interlayer exchange (shown in blue in Fig. 3f). As in any exchange process, geometrical parameters that influence the relevant hopping integrals play a major role. In superexchange, the angle between magnetic ions and its ligands mediates the superexchange, in this case the Mn-Te-Mn bond angle $\theta$ shown in Fig. 3f. These structural superexchange interactions are further controlled by orbital hybridization with cationic Bi $p$ states tuned by the nearest-neighbor ions across the vdW gap[20], in this case, determined by the Te-Te distance $\Delta$ shown in Fig. 3f, to stabilize the FM interlayer coupling in MnBi$_2$Te$_4$.

We first note that $A_{1g}$ modes in MnBi$_2$Te$_4$ modulate $\Delta$, whereas $E_g$ modes do not, an observation that accounts for the dichotomy of their respective influence on $J^\perp$. Of the $A_{1g}$ modes, examining the eigenvectors in Fig. 1b and Fig. S1b, $A_{1g}^{(2)}$ exhibits the largest modulation of the Mn-Te-Mn bond angle $\theta$. The modulation of $\theta$ by the $A_{1g}^{(2)}$ mode is a factor of 2 larger than by $A_{1g}^{(1)}$, which in turn is a factor of 5 larger than by $A_{1g}^{(3)}$. This rationalizes the trend seen in the calculated $\frac{dJ^\perp}{du}$ in terms of the SSE pathways.

Finally, we investigate magnetophononic coupling by direct measurement of phonons in the time domain. To do this, we carry out 'pump-probe' experiments to generate and detect coherent optical phonons as a function of magnetic field (see schematic in Fig. 4a). Excitation with ultrafast optical pump pulses (1.55 eV, 50 fs) results in the generation of coherent phonon oscillations. A second, time-delayed probe pulse (1.2 eV, 50 fs) measures pump-induced changes in the transient reflectivity ($\Delta R/R$). The transient reflectivity is sensitive to changes in carrier density and coherent phonons. These measurements are carried out at 2 K, as a function of magnetic field from 0 to 6.4 T, across the spin-flop transition. The transient reflectivity, shown in Fig. 4b, exhibits an initial sub-picosecond dip, followed by a slow relaxation. Overlayed on this are multiple distinct coherent oscillation components that (as described below), correspond to the $A_{1g}^{(1)}$ and $A_{1g}^{(2)}$ phonons. We normalize the pump-probe reflectivity traces with respect to their maximum amplitudes, to account for field-dependent variation in the absorbed fluence, and thus photocarrier density, which can influence coherent phonon amplitudes (see SI Section S7 and Fig. S8 for detailed discussion). Upon subtracting bi-exponential fits (black line fit to 0 T data in Fig. 4b shown as a representative example), we observe that the normalized phonon oscillation amplitudes in the residual $\Delta R/R$ in Fig. 4b visibly decrease with increasing magnetic field, much like the phonon spectral weights measured using Raman spectroscopy.

The individual oscillatory components are obtained by fitting the residual $\Delta R/R$ to the sum of two exponentially decaying sinusoidal functions (see Methods) as shown for the representative 0 T data in Fig. 4c (see SI for fitting of full dataset). The individual sinusoidal functions, shown in Fig. 4d, are readily identified as the $A_{1g}^{(1)}$ and $A_{1g}^{(2)}$ modes at 1.47 THz (49 cm$^{-1}$) and 3.44 THz (115 cm$^{-1}$), respectively. Plotting the amplitudes of the two coherent phonon modes as a function of magnetic field in Fig. 4e, it is clear that both modes track the AFM order parameter denoted by the solid grey line, in striking similarity to the field-dependent change in the Raman scattering intensities.

The detection of coherent phonons in pump-probe experiments occurs through a process that is identical to spontaneous Raman scattering[21,22]. The generation of coherent phonons can also be described within a Raman formalism, with the real and imaginary parts of the Raman tensor responsible for phonon excitation in transparent and absorbing materials, respectively[21]. The similarity of the magnetic-field dependent coherent phonon amplitudes in Fig. 4e to the static Raman scattering intensities in Fig. 2b thus suggests that these are a consequence of the same mechanism, namely the excitation of zone-boundary phonons via the crystal momentum associated with the antiferromagnetic order.

For resonant excitation of MnBi$_2$Te$_4$ with 1.55 eV pulses, phonon excitation through the imaginary part of the Raman tensor may be physically thought of in terms of a 'displacive' excitation[23], where the ultrafast excitation of carriers by the pump pulse shifts the quasi-equilibrium coordinates of the lattice in a spatially and temporally coherent manner, generating coherent phonons. Within this picture, magnetophononic zone-folding as described in the previous section would allow for the generation of both zone-center as well as nominally zone-boundary $A_{1g}$ modes. Additionally, the electronic excitation that shifts the quasi-equilibrium coordinates may itself have a $q_z = \pi/c$ component owing to the contrast in spin-split electronic bands in alternating layers, acting as a direct driving force for the generation of zone-boundary phonons. Unfortunately, the small frequency splitting of the $A_{1g}$ modes precludes the explicit resolution of zone-boundary phonons in the time domain. Nonetheless, it is clear from Fig. 4d that the coherent phonons track the AFM order parameter in accord with the magnetophononic wavemixing proposed here.

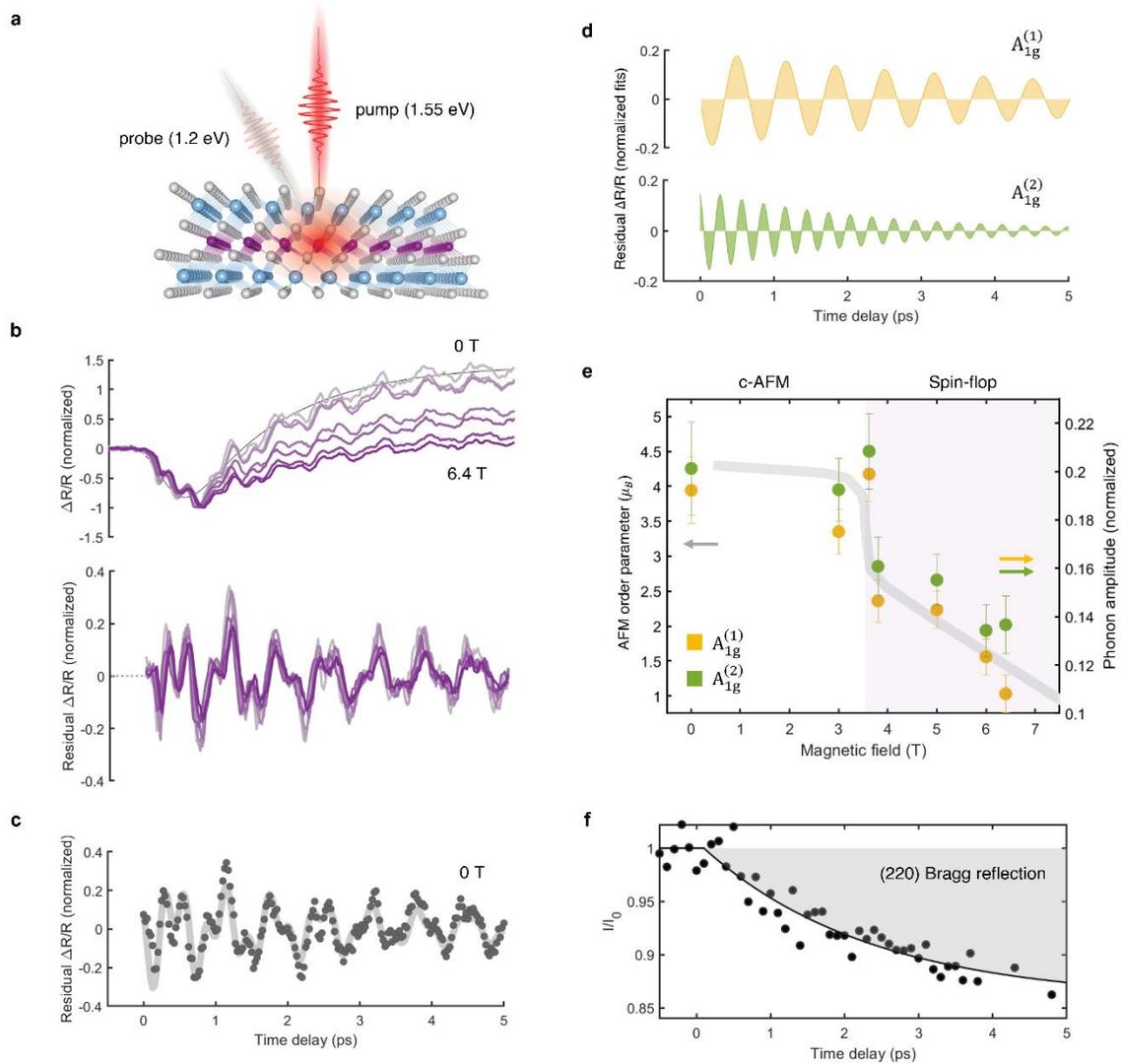

**Fig. 4. Ultrafast signatures of magnetophononic coupling. a,** Schematic of pump-probe measurement. **b,** Pump-induced changes in the transient reflectivity ($\Delta R/R$) as a function of time delay at various magnetic fields, normalized to the maximum amplitude. The black line is a representative biexponential fit to the 0 T data. The lower panel shows the residual $\Delta R/R$ upon subtracting a biexponential fit. **c,** Residual $\Delta R/R$ at 0 T, with black dots denoting experimental datapoints and the grey line denoting the fit to the sum of two decaying sinusoidal functions. **d,** Individual decaying sinusoidal components obtained from the fit in panel b, corresponding to the $A_{1g}^{(1)}$ (top) and $A_{1g}^{(2)}$ (bottom) phonons, respectively. **e,** Initial amplitude of the coherent $A_{1g}^{(1)}$ and $A_{1g}^{(2)}$ phonons, obtained from fit result in panel c. The grey line is the antiferromagnetic order parameter from reference[15]. **f,** Measured transient electron diffraction intensity of the (2 2 0) Bragg peak, with black dots denoting experimental datapoints, and the black line denoting the fit to an exponential decay function. Error bars are standard deviations in fit values.

We note that in general, phonons in time domain measurements are expected to exhibit qualitative deviations from steady-state spectroscopy, owing to the nonequilibrium nature of the former. While the ultrafast carrier excitation in displacive phonon excitation is itself a manifestly nonequilibrium process, additional deviations may emerge from nonequilibrium phonon interactions. We directly measure the timescale of phonon equilibration using ultrafast electron diffraction (see Methods). Here, pump-induced changes in the root-mean-square displacements $\langle u^2 \rangle$ of ions through carrier-lattice and lattice thermalization appear in the transient intensity of Bragg peaks through the Debye-Waller effect (see SI Section S6). As a representative sample, we show in Fig. 4f, the transient intensity of the (2 2 0) Bragg peak, with the evolution of the peak intensity fit to an exponential decay (black line). The results indicate that phonon populations indeed remain in a nonequilibrium state through the entire time delay range considered. It is noteworthy that clear signatures of magnetophononic coupling are observed even under such nonequilibrium conditions. Finally, we mention that there may possibly be additional contributions to the coherent phonon amplitudes from magnetodielectric effects which are not explicitly accounted for here. We discuss the possible contributions to coherent phonon amplitudes due to such an effect Supplementary Section S7.

**Discussion**

We have demonstrated that optically 'forbidden' zone-boundary phonons are observed due to magnetophononic wave-mixing in MnBi$_2$Te$_4$. While it is uncommon for purely magnetic unit cell doubling to give rise to phonon zone-folding effects, such signatures were first observed in transition metal dihalides[24]. These observations were rationalized in terms of phenomenological models of electron-phonon coupling that took into consideration phonon modulation of the spin-orbit coupling and exchange interactions[24,25]. Our model instead considers the scattering cross-section between the AFM order and phonons, arriving at qualitatively similar conclusions. Importantly, our work provides a description of such a model using first-principles theory. The excellent agreement between the theory and experimental results not only validates the model, but also provides a microscopic basis for the observed phenomena in terms of SSE interlayer exchange pathways. Our work may also help rationalize similar phenomena recently reported[26,27] in other quasi-two-dimensional magnets such as CrI$_3$ and FePS$_3$.

Our discovery is especially of significance in light of the critical role played by tunable interlayer exchange interactions in layered magnetic materials. For instance, in MnBi$_2$Te$_4$, the interlayer magnetic ordering can drive topological phase transitions between quantum anomalous Hall and axion insulator states. Our work unlocks the possibility of controlling the interlayer magnetic ordering in MnBi$_2$Te$_4$ by exploiting the strong coupling of $A_{1g}$ phonons to $J^\perp$. A promising route towards the ultrafast control of magnetism in MnBi$_2$Te$_4$ is the use of resonant THz excitation to drive large amplitude distortions along $A_{1g}$ modes, as opposed to employing carrier-based mechanisms (such as displacive excitation) that suffer from ultrafast heating effects, which limit the amplitude of coherent phonons. This may be through anharmonic coupling to Raman active modes[28], or alternatively through sum-frequency ionic Raman scattering[29]. Such mechanisms based on resonant coupling have been used to drive ultrafast light-induced magnetic oscillations and phase transitions, as experimentally demonstrated in other materials[30–36]. Experimental studies[37] on Bi$_2$Se$_3$, a material closely related to MnBi$_2$Te$_4$, have demonstrated the feasibility of ionic Raman scattering as a way to drive large amplitude oscillations along Raman active modes. Recent theoretical work[38] has outlined an approach based on anharmonic phonon interactions in MnBi$_2$Te$_4$. In particular, it was shown that resonant excitation of an IR-active $A_{2u}$ phonon (at a frequency of 156 cm$^{-1}$ = 4.7 THz) could drive large amplitude oscillations, which via anharmonic coupling, would drive a unidirectional distortion along Raman-active $A_{1g}$ modes such as the ones identified in the present work. It was predicted that such an approach could be used to drive an AFM to FM transition concurrent with a topological phase transition, using experimentally accessible ultrafast modalities. The magnetophononic wave-mixing in the present work provides an experimental foundation for such approaches and a path towards achieving ultrafast light-induced topological phase transitions.

## Methods

**Crystal growth and characterization**

Single crystals of MnBi$_2$Te$_4$ were grown using a self-flux method as reported elsewhere[12]. The phase and crystallinity of the single crystals were checked by X-ray diffraction. The antiferromagnetic order with the Néel temperature of 24 K was confirmed using SQUID magnetometry.

**Raman spectroscopy measurements**

Temperature-dependent Raman spectra were collected using a Horiba LabRam HR Evolution with a freespace Olympus BX51 confocal microscope. A 632.8 nm linearly polarized HeNe laser beam was focused at normal incidence using a LWD 50x objective with a numerical aperture of 0.5, with the confocal hole set to 100 $\mu$m. A Si back-illuminated deep depleted array detector and an ultra-low-frequency volume Bragg filter were used to collect the spectra, dispersed by a grating (1800 gr/mm) with an 800 mm focal length spectrometer. The system was interfaced with an Oxford continuous-flow cryostat for low-temperature measurements, using liquid helium as the cryogen.

Field-dependent magneto-Raman spectra were collected using a home-built Raman spectrometer. A 632.8 nm linearly polarized HeNe laser beam was focused at normal incidence using a LWD 50x objective with a numerical aperture of 0.82. A Si back-illuminated deep depleted array detector and a set of ultra-low-frequency volume Bragg filters were used to collect the spectra, dispersed by a grating (1800 gr/mm) with a 300 mm focal length spectrometer. The system was interfaced with an Attocube AttoDRY 2100 closed-cycle cryostat for low-temperature, high magnetic field measurements, using liquid Helium as the cryogen. The field-induced Faraday rotation in the objective was calibrated and corrected using a half-waveplate.

The laser power was maintained below 50 $\mu$W in all measurements, to minimize laser heating and maintain the power well below the damage threshold. Laser heating was calibrated by measuring Raman phonon peak shifts as a function of and using thermal conductivity values from reference[14]. Polarized spectra were obtained using a half-waveplate to rotate the polarization of the incident beam, with a fixed analyzer.

After peak assignment using polarization analysis, temperature- and field-dependent spectra were collected without a polarizer, to maximize signal throughput. Spectra were averaged over 60 minutes and 120 minutes in the case of temperature-dependent and field-dependent measurements respectively, with a temperature stability of $\pm 0.1$ K. Any subtle drift in the spectrometer (<0.15 cm$^{-1}$) over the temperature-dependent studies was corrected using the HeNe line at 632.8 nm.

The $A_{1g}^{(1)}$ peak was fit using an inverse Fano lineshape in combination with a linear background. Its lineshape is given by the expression $I(\omega) = \frac{(q\Gamma - (\omega - \omega_0))^2}{\Gamma^2 + (\omega - \omega_0)^2}$, where $I$ is the scattering intensity, $\omega$ is the energy, $\omega_0$ and $\Gamma$ are the resonant energy and linewidth of the excitation respectively, and $1/q$ is a measure of the peak asymmetry. The $E_g^{(2)}$, and $A_{1g}^{(3)}$ peaks were fit with a standard Gaussian lineshape, and the $E_{1g}^{(3)}$ and $A_{1g}^{(2)}$ peaks were fit with a standard Lorentzian lineshape.

A nonlinear least-squares fitting procedure was used. To ensure robustness of the temperature-dependent fits, the same initial fit values and constraints were used for each set of temperature-dependent and field-dependent spectra.

**Magnetic field-dependent ultrafast optical spectroscopy**

Ultrafast optical pump-probe measurements were carried out using a 1040 nm 200 kHz Spectra-Physics Spirit Yb-based hybrid-fiber laser coupled to a noncollinear optical parametric amplifier. The amplifier produces <50 fs pulses centered at 800 nm (1.55 eV), which is used as the pump beam. The 1040 nm (1.2 eV) output is converted to white light, centered at 1025 nm with a FWHM of 20 nm, by focusing it inside a YAG (Yttrium Aluminum Garnet) crystal. The white light is subsequently compressed to ~50 fs pulses using a prism compressor pair and is used as the probe beam. The pump and the probe beams are aligned to propagate along the [001] axis of the crystal, at near normal incidence.

The samples were placed in a magneto-optical closed-cycle cryostat (Quantum Design OptiCool). Pump-probe measurements were carried out as a function of magnetic field applied normal to the sample surface (along the [001] direction). The sample temperature was fixed at 2 K. A pump fluence of ~100 µJ/cm$^2$ was used in order to generate sufficiently large coherent phonon oscillations, while keeping the transient heating to a minimal amount, to ensure we avoid melting of the magnetic order.

**Ultrafast electron diffraction measurements**

Ultrafast electron diffraction measurements were carried out at the MeV-UED beamline at the SLAC National Accelerator Laboratory. The principle and other technical details of the experimental setup are outlined elsewhere[39]. A 60-fs laser pulse with a photon energy of 1.55 eV and fluence of 7 mJ/cm$^2$ were used to excite the sample. A higher pump fluence was required than in the optical pump-probe measurements, in order to produce a sufficiently large pump-induced change in diffraction intensities. Fluence-dependent damage studies revealed no signs of laser-induced damage, and the measurements were repeatable over thousands of cycles. Femtosecond electron bunches of ~100 fs pulsewidth and 3.7 MeV kinetic energy were used to measure pump-induced changes in electron diffraction intensities.

Measurements were carried out on flakes with an average thickness of around 100 nm, exfoliated from a single crystal of MnBi$_2$Te$_4$ and transferred onto an amorphous Si$_3$N$_4$ membrane using an ex-situ transfer stage. The flakes were protected with an additional layer of amorphous Si$_3$N$_4$ to prevent degradation. The spot sizes of the pump and probe beams were 464×694 µm and ~70 µm, respectively, and the measurements were carried out at 30 K.

The ultrafast electron diffraction intensities were obtained by averaging over several scans, normalizing individual diffraction images to account for electron beam intensity fluctuation. Individual diffraction peaks were fit to a two-dimensional Gaussian function, and then averaged over symmetry-related peaks based on the *R*-3*m* space group of MnBi$_2$Te$_4$.

**Pump-probe data analysis**

The time-resolved reflectivity traces were first fitted to a product of an error function and a biexponential decay function. The error function models the excitation of photo-carriers and instrumental temporal resolution, and the exponential decay is an approximation for the sum of various unknown processes occurring over the measured time delay, including electron-electron and electron-phonon thermalization. The functional form is -

$$\left(1 + \mathrm{erf}\left(\frac{t}{\tau_r}\right)\right) \times \left(A_1 \exp\left(-\frac{t}{\tau_1}\right) + A_2 \exp\left(-\frac{t}{\tau_2}\right) + C\right),$$

where $t$ is the time delay, $\tau_{el}$ is the rise time for the excitation of photo-carriers, $\tau_1$ and $\tau_2$ are the time constants of exponential decay, and $A_1$, $A_2$, and $C$ are constants.

Upon subtracting the biexponential decay, the residual traces were fit to the sum of two decaying sinusoidal functions. The functional form is –

$$A_1 \sin(2\pi f_1 t + \phi_1) \exp\left(-\frac{t}{\tau_{d1}}\right) + A_2 \sin(2\pi f_2 t + \phi_2) \exp\left(-\frac{t}{\tau_{d2}}\right),$$

where $t$ is the time delay, $f_1$ and $f_2$ are the frequencies of the sinusoidal functionals, corresponding to the $A_{1g}^{(1)}$ and $A_{1g}^{(2)}$ phonons, $\phi_1$ and $\phi_2$ are the phases, and $\tau_{d1}$ and $\tau_{d2}$ are the time constants of exponential decay of the oscillations. The initial amplitudes $A_1$ and $A_2$ are plotted in Fig. 4d.

The ultrafast electron diffraction intensities were fit to an exponential decay function of the form –

$$A_1 \exp\left(-\frac{t}{\tau_l}\right) + C,$$

where $t$ is the time delay, $\tau_l$ is the time constant, and $A_1$ and $C$ are constants.

**Electronic structure and phonon calculations**

Density functional theory calculations were carried out using the Vienna Ab Initio Simulation Package (VASP)[40–44] with the PBE exchange correlation functional[45] and van der Waals correction via the DFT-D3[46,47] method with Becke-Jonson damping. A Hubbard $U$ was also added to the Mn (4 eV) using Dudarev's[48] approach. A non-primitive cell containing two Mn atoms was used to obtain the equilibrium geometry of the system with AFM-A magnetic structure. Γ-point phonons were obtained with the finite displacement method on a 1×1×1 'supercell' using the PHONOPY software package[49] and VASP. An energy cutoff of 300 eV was used for all calculations. A 4×4×4 Γ-centered $k$-point mesh was used for equilibrium relaxations and phonon calculations. The general energy convergence threshold was $1\times10^{-8}$ eV and the force convergence threshold for relaxation was $1\times10^{-5}$ eV/Å. When including SOC in the magnetic parameter calculations, however, the energy convergence threshold was $1\times10^{-6}$ eV. Gaussian smearing with a 0.02 eV width was also used in all relaxation and single-point energy calculations. Density of states calculations employed the tetrahedron method. The metallic state was modelled by electron doping the unit cells with 0.1 electron/Mn atom. Supercells for magnetic exchange calculations were generated using VESTA[50].

**Exchange coupling constants calculations**

Magnetic exchange parameters were obtained by considering a model spin Hamiltonian of the form $H = -\sum_{\langle ij \rangle} J_{ij} S_i \cdot S_j$, where $J_{ij}$ includes intra-layer exchange parameters $J_1$ and $J_2$ and inter-layer exchange parameter $J^\perp$. A $\sqrt{2}\times\sqrt{2}\times1$ supercell of the conventional cell was used get the intra-layer exchange parameters, while a 1×1×2 supercell of the primitive cell was used to get the inter-layer exchange parameter. Γ-centered $k$-point meshes of 4×4×1 and 4×4×4 were used in the respective calculations.

For the intra-layer exchange parameters, one FM and two AFM configurations (stripe and up-up-down-down) were used. The spin exchange energy equations in terms of magnetic exchange parameters for structures of $R\bar{3}m$ symmetry are as follows:

$$E_{FM} = 3E_{NM} - 60J_1 S_i \cdot S_j - 60J_2 S_i \cdot S_j$$

$$E_{AFM1} = 3E_{NM} + 12J_1 S_i \cdot S_j + 12J_2 S_i \cdot S_j$$

$$E_{AFM2} = 3E_{NM} + 12J_1 S_i \cdot S_j - 12J_2 S_i \cdot S_j$$

For the inter-layer exchange parameter, one FM and one AFM configuration were used.

$$E_{FM} = E_{NM} - 6J^\perp S_i \cdot S_j$$

$$E_{AFM-A} = E_{NM} + 6J^\perp S_i \cdot S_j$$

The calculated values were multiplied by $S^2$ to obtain the exchange coupling in meV, assuming the spin of the local moment is $S = 5/2$.

**Acknowledgements.** H.P., V.A.S., H.W., P.K., M.P., N.Z.K., A.M.L., R.A., J.M.R., and V.G. acknowledge support from the DOE-BES grant DE-SC0012375. H.P. acknowledges partial support from the DOE Computational Materials program, DE-SC0020145. Support for crystal growth and characterization was provided by the National Science Foundation through the Penn State 2D Crystal Consortium-Materials Innovation Platform (2DCC-MIP) under NSF cooperative agreement DMR-1539916. D.P. was supported by the Army Research Office (ARO) under grant no. W911NF-15-1-0017. SLAC MeV-UED is supported in part by the DOE BES SUF Division Accelerator & Detector R&D program, the LCLS Facility, and SLAC under Contract Nos. DE-AC02–05-CH11231 and DE-AC02–76SF00515.

**Author contributions.** H.P. and V.G. conceived the project. Raman spectroscopy measurements and analysis were carried out by H.P., H.W., M.W., and V.G. Pump-probe reflectivity was carried out by P.K., M.P., H.P., R.S., and R.A., and the results analyzed by H.P., P.K., M.P., and V.A.S. Density functional theory calculations were done by N.Z.K., D.P., M.G., and J.M.R. Ultrafast electron diffraction was carried out by H.P., V.A.S., H. W., X.S., A.H.R., A.M.L., and X.W., and the results were analyzed by H.P. Crystal growth and characterization were done by S.H.L. and Z.Q.M. The paper was written by H.P. with inputs from all authors.

**Competing interest declaration.** The authors declare no competing interests.

**Additional information.** Supplementary Information is available for this paper. Correspondence and requests for materials should be addressed to H.P., R.A., and V.G.

## S1. Raman peak assignment and eigenvectors

We start with a systematic analysis of Raman phonon spectra, shown in Fig. S1a. The non-magnetic unit cell contains seven atoms, and thus there are 21 phonon modes in total, consisting of 18 optical and 3 acoustic modes. Using representation theory, these can be decomposed into irreps of the point group $\bar{3}m$. Of these, only the $E_g$ and $A_{1g}$ modes are Raman active. Polarized Raman spectroscopy measurements are used to readily identify these modes based on their different selection rules. In particular, $E_g$ modes have non-vanishing diagonal Raman tensor components and are thus visible under both parallel- and cross-polarized configurations, whereas the $A_{1g}$ modes have only diagonal Raman tensor components and are visible only under the parallel-polarized configuration. We did not observe any dependence on the in-plane crystallographic orientation.

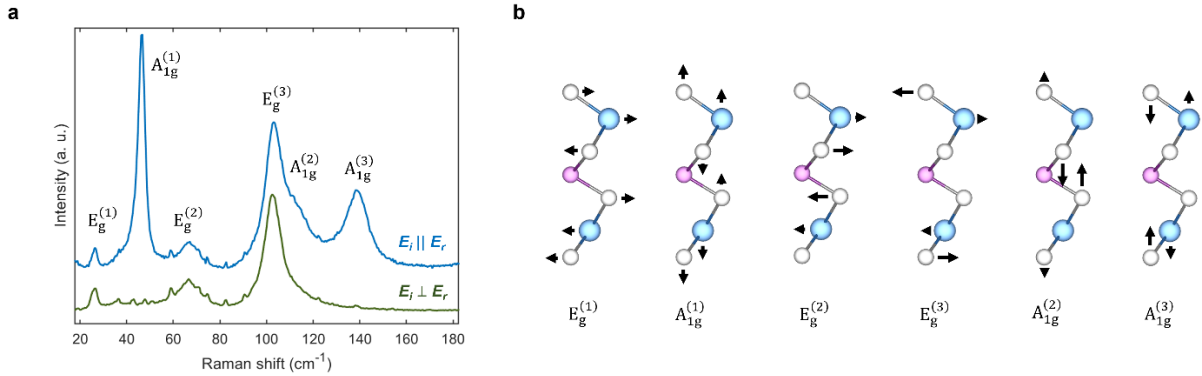

**Fig. S1 Polarized Raman spectra. a,** Raman spectra with the incident and reflected beams parallel- and cross-polarized with respect to each other, at 298 K. **b,** Eigenvectors of Raman phonons, with the arrows denoting ionic motions, calculated using density functional theory simulations. The arrow lengths are proportional to the actual calculated ionic eigendisplacements for all modes.

First-principles calculations are used to enumerate all the Γ-point Raman-active optical phonon modes and their energies in Table S1. Good agreement is obtained between the calculations and measurements for all the observed Raman phonons, confirming that the first-principles calculations provide a good description of the lattice dynamics.

| Symmetry | Frequency (cm$^{-1}$) (theory, DFT) | Frequency (cm$^{-1}$) (experiment at 15 K) |
| --- | --- | --- |
| $E_g$ | 32.9 | 27.2 |
| $A_{1g}$ | 50.7 | 49.1 |
| $E_g$ | 79.8 | 69.8 |
| $E_g$ | 112.9 | 108.3 |
| $A_{1g}$ | 119.2 | 113.1 |
| $A_{1g}$ | 148.9 | 146.6 |

**Table. S1 | Raman phonon mode assignment.** Raman phonon symmetries and frequencies at the Γ point, from density functional theory calculations (theory), and Raman spectroscopy (experiment) at 15 K.

The phonon eigendisplacements at the Γ point, calculated using density functional theory simulations, show that $A_{1g}$ phonons have purely out-of-plane ionic motions, whereas $E_g$ phonons have purely in-plane ionic motions.

## S2. Anomalous temperature-dependence of $A_{1g}^{(1)}$ mode

In Fig. S2a, we plot the Raman spectra measured at 15 K and 300 K, normalized to the height of the $E_g^{(3)}$ peak at ~113 cm$^{-1}$, for convenience. We note that the result identified below is independent of the choice of normalization. In general, phonon peaks in Raman spectra broaden with increasing temperature due to increased phonon-phonon scattering, with resultant lower peak heights. This is visible for instance in the $A_{1g}^{(3)}$ peak at ~145 cm$^{-1}$. On the other hand, the scattering intensity of the $A_{1g}^{(1)}$ mode exhibits an anomalous temperature-dependence, with a dramatic *decrease* in height and integrated intensity, with decreasing temperature. It is apparent that this decrease in amplitude is independent of the choice of normalization. The amplitude does not show any clear correlation with the magnetic transition at $T_N = 24$ K.

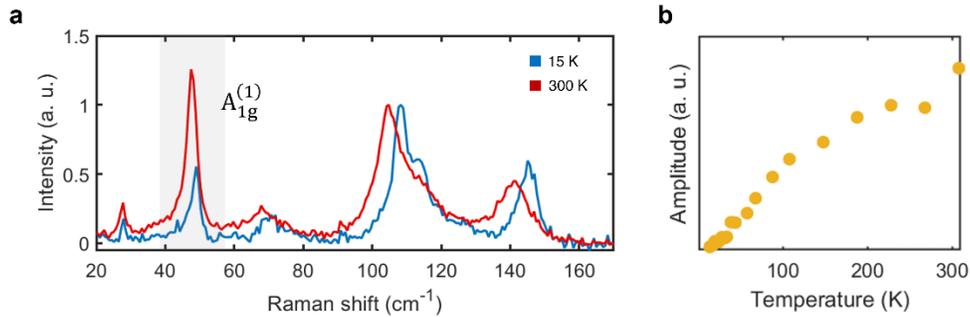

**Fig. S2 Anomalous temperature-dependence of $A_{1g}^{(1)}$ amplitude. a,** Unpolarized Raman spectra at 15 K and 300 K, normalized to the height of the $E_g^{(3)}$ peak at ~115 cm$^{-1}$. The $A_{1g}^{(1)}$ mode is highlighted in grey. **b,** The amplitude of the $A_{1g}^{(1)}$ peak, fit to a Fano lineshape, as outlined in the Methods section.

A possible explanation for the dramatic change in scattering intensity with temperature is proximity of the Raman excitation energy (633 nm = 1.96 eV) to electronic transitions correlated with the ionic motion of the $A_{1g}^{(1)}$ mode. Optical conductivity measurements[1] indeed show large changes in the measured temperature range. Such resonant effects may be probed by measuring relative Raman phonon scattering cross-sections as a function of the excitation energy. Resonant effects are discussed in detail in Section S5.

The scattering intensity associated with the zone-boundary $A_{1g}^{(1)}$ mode is clearly visible in the field-dependent Raman spectra in Fig. 1d in the main text. In the temperature-dependent Raman spectra in Fig. 1c however, the anomalous temperature-dependence, described above, appears to swamp the small zone-boundary scattering intensity.

# S3. Field-dependence of $E_g^{(2)}$, $E_g^{(3)}$, and $A_{1g}^{(3)}$ spectral weights

The integrated intensities of the $E_g^{(2)}$, $E_g^{(3)}$, and $A_{1g}^{(3)}$ phonons are plotted in Fig. S3 as a function of magnetic field. The integrated intensities were obtained by fitting individual spectra following the procedure outlined in the Methods section, with the error bars denoting the standard deviation in fit values. We note a small dip in the $E_g^{(2)}$ intensity at the spin-flop critical field of 3.7 T. Outside of this, the three modes shown here exhibit no clear field-dependent behavior above the experimental and fitting uncertainty. In particular, there is no signature of coupling to the antiferromagnetic order parameter and the associated zone-boundary phonons.

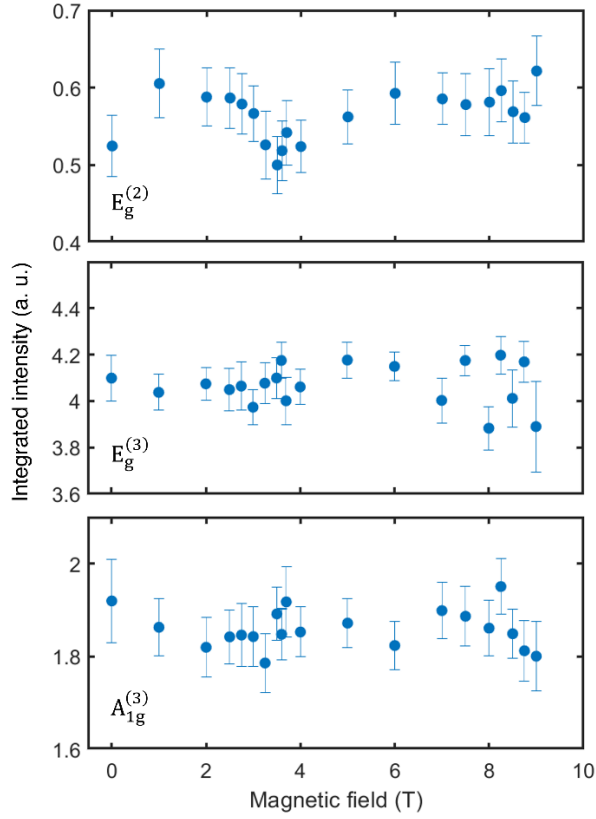

**Fig. S3 Absence of magnetophononic coupling.** The panels show spectral weights of the $E_g^{(2)}$, $E_g^{(3)}$, and $A_{1g}^{(3)}$ modes respectively, as a function of magnetic field. The error bars are standard deviations of the fit values.

## S4. Generalized magnetophononic coupling and frequency renormalization

We write down minimal lattice and spin Hamiltonians[2] to describe a generalized magnetophononic coupling. Consider the lattice Hamiltonian described by the harmonic approximation,

$$H_L = H_L^0 + \frac{1}{2!}\frac{\partial^2 H_L^0}{\partial u_\alpha^2}u_\alpha^2 + O(u_\alpha^3) \approx H_L^0 + \frac{1}{2}N\mu_\alpha {v_\alpha^0}^2 u_\alpha^2, \quad (1)$$

where $u_\alpha$ is the displacement along the phonon normal mode $\alpha$, $\mu_\alpha$ is the reduced mass, $v_\alpha$ is the frequency, and $N$ is the number of unit cells. The magnetic ground state energy described by a Heisenberg-like Hamiltonian,

$$H_M^0 = -\sum_{ij} J_{ij} S_i \cdot S_j, \quad (2)$$

where $i$ and $j$ are spin site indices, and $J_{ij}$ is the isotropic exchange interaction between spins at $i$ and $j$. When this is perturbed by a zone-center optical phonon $\alpha$, the perturbed magnetic energy can be derived by considering the derivatives of $J_{ij}$ with respect to the phonon normal mode displacement $u_\alpha$. Expanding upto second order in $u_\alpha$, the perturbed exchange interaction is

$$J'_{ij}(u_\alpha) = J_{ij} + \frac{\partial J_{ij}}{\partial u_\alpha}u_\alpha + \frac{1}{2}\frac{\partial^2 J_{ij}}{\partial u_\alpha^2}u_\alpha^2. \quad (3)$$

Here, the first order term, in the specific case of $J = J^\perp$ is responsible for the magnetophononic wave-mixing described in detail in the main text (Eq. 1 and Eq. 2).

The second order term, proportional to $u_\alpha^2$, renormalizes the harmonic term in the lattice energy, resulting in spin-induced phonon frequency changes. Separating the in-plane and out-of-plane exchange couplings, denoted by $J_\mu$ and $J_\mu^\perp$, respectively, where $\mu = 1, 2, \ldots$ are the first- and second-nearest-neighbors and so on, and assuming small spin-induced energy shifts i. e. $v_\alpha + v_{\alpha 0} \approx 2v_{\alpha 0}$, the renormalized phonon frequency is given by

$$v_\alpha - v_{\alpha 0} = \frac{1}{4N\mu_\alpha v_{\alpha 0}}\left[\sum_\mu \frac{\partial^2 J_\mu}{\partial u_\alpha^2}\sum_i S_i \cdot S_{i+\mu} + \sum_\mu \frac{\partial^2 J_\mu^\perp}{\partial u_\alpha^2}\sum_i S_i \cdot S_{i+\mu^\perp}\right]. \quad (4)$$

The above expression shows the renormalization of the phonon frequency due to spin order along different directions, through the respective exchange couplings. Under a mean-field approximation, (4) simplifies to $v_\alpha - v_{\alpha 0} \propto \langle S^2 \rangle$.

Experimentally, we observe such a spin-induced phonon frequency renormalization in the $A_{1g}^{(1)}$ mode. The phonon frequencies are first extracted as a function of temperature, using the fitting procedure outlined in the Methods section of the main text. We then account for phonon-phonon interactions by fitting the temperature-dependent phonon frequencies to that of a (cubic) anharmonic phonon, given by $\omega(T) = \omega_0 + A\left(1 + \frac{2}{e^{\frac{\hbar\omega_0}{2k_BT}}-1}\right)$, where $\omega$ is the phonon frequency renormalized by anharmonic (phonon-phonon) interactions, $T$ is the temperature, $\omega_0$ is the bare phonon frequency, and $A$ is a mode-specific fitting constant.

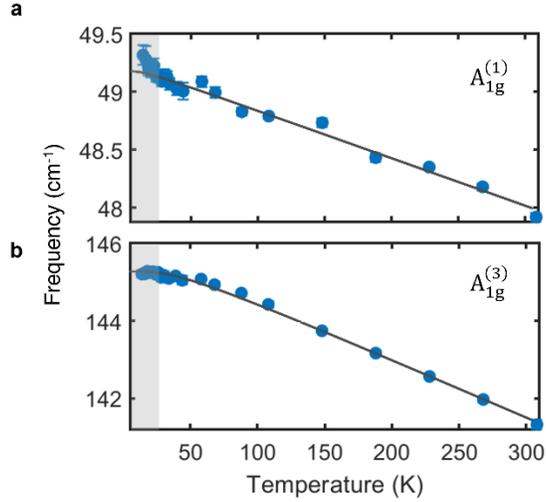

**Fig. S4 Spin-induced phonon frequency renormalization.** Temperature-dependent frequency of the $A_{1g}^{(1)}$ phonon mode (top) and $A_{1g}^{(3)}$ phonon mode (bottom). The black lines are fits to the anharmonic phonon model described in the text.

Fig. S4 shows the temperature-dependent frequency of the $A_{1g}^{(1)}$ mode, with the black line showing a fit to the anharmonic phonon model. The plot shows a small but clear deviation from the fit below $T_N = 24$ K, indicating a spin-induced phonon frequency renormalization. In contrast, the temperature-dependent frequency of the $A_{1g}^{(3)}$ mode in Fig. S3 shows good agreement with the anharmonic phonon model down to the lowest temperatures, indicating that the $A_{1g}^{(3)}$ mode does not exhibit a significant spin-induced frequency renormalization.

Interestingly, we note that a previous study[3] on atomically thin flakes of $MnBi_2Te_4$ reported a negative spin-induced frequency renormalization of the $A_{1g}^{(1)}$ mode, contrary to the positive frequency renormalization observed in the bulk crystals used in our study. This difference may possibly be due to changes in the electronic and magnetic structure as a function of sample thickness in the 2D limit.

The strong magnetophononic coupling observed in the $A_{1g}^{(2)}$ mode in our magneto-Raman measurements suggests that it too might exhibit a significant spin-induced frequency shift. Unfortunately, the spectral overlap between the $A_{1g}^{(2)}$ and $E_g^{(3)}$ modes (see Fig. S1a) and strong $A_{1g}^{(2)}$ zone-boundary scattering intensity below $T_N$ hinders a similar temperature-dependent frequency analysis for the $A_{1g}^{(2)}$ mode. None of the other observed Raman phonons exhibit a spin-induced frequency renormalization above the experimental uncertainty.

## S5. Resonant Raman effects

Resonant Raman effects may potentially give rise to temperature- and field-dependent artifacts in phonon peak intensities due to changes in the electronic band structure across phase transitions. In order to rule out such an explanation for the phenomena reported in Figs. 1 and 2, we investigate resonant Raman effects in MnBi$_2$Te$_4$ by measuring phonon spectra at different laser excitation energies. In Fig. S6, we show the Raman phonon spectra measured with laser excitation energies of 1.58 eV (785 nm), 1.96 eV (633 nm), and 2.71 eV (458 nm), at 297 K and zero magnetic field. Note that the 1.58 eV Raman spectrum is only shown down to 65 cm$^{-1}$ due to the limitation of our low-frequency filters at this excitation wavelength.

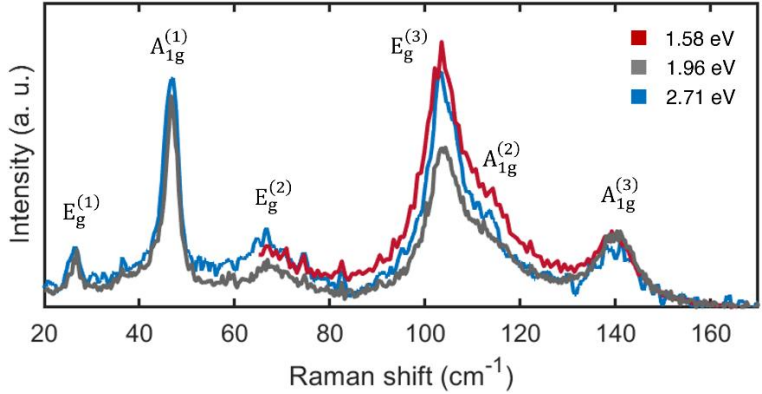

**Fig. S5 Phonon spectra at different laser excitation energies.** Raman phonon spectra measured using 1.58 eV (785 nm), 1.96 eV (633 nm), and 2.71 eV (458 nm) laser excitation energies at 297 K and zero magnetic field. Spectra are normalized to the $A_{1g}^{(3)}$ peak intensity.

It is observed that phonon peak intensities indeed change as a function of the laser excitation energy, however, these changes occur across all the observed phonon modes, i. e. the three $A_{1g}$ modes as well as the three $E_g$ modes. It is clear that this result is independent of the choice of normalization. In contrast, the temperature- and field-dependent magnetophononic effects observed in our study are only in the $A_{1g}^{(1)}$ and $A_{1g}^{(2)}$ modes, with negligible changes in the scattering intensities of other modes. Our observations reported in Figs. 1 and 2 are thus inconsistent with resonant Raman effects.

Furthermore, upon tracking the $A_{1g}^{(2)}$ mode across the PM → AFM transition at 24 K using the 1.58 eV (785 nm) laser excitation, it is found that the Raman scattering intensity exhibits quantitatively the same behavior (see Fig. S6) as with the 1.96 eV (633 nm) laser excitation (see Fig. 1c) – i. e. the $A_{1g}^{(2)}$ scattering intensity is enhanced by around 35% in the AFM phase. This is additional evidence that the observed phenomenon is inconsistent with resonant Raman effects, wherein different excitation energies would give rise to qualitatively different temperature-dependent intensity changes. It is instead consistent with an effect arising from the AFM order, as in our model of magnetophononic wavemixing.

It is useful to consider the exchange energies involved in various magnetic phase transitions in MnBi$_2$Te$_4$. The dominant in-plane nearest neighbor exchange coupling is 0.12 meV, whereas the interplanar exchange coupling is an order-of-magnitude weaker[4]. The temperature-driven PM → AFM transition is accompanied by significant magnetic energy changes due to the in-plane ordering of spins, and the large in-plane exchange coupling. On the other hand, the in-plane ordering remains unchanged in the out-of-plane magnetic field-driven AFM → FM transition, with only the interplanar magnetic order being modulated. The accompanying magnetic energy changes are thus an order of magnitude weaker than in the PM → AFM transition. Hence it is expected that the associated electronic structure changes as a function of out-of-plane magnetic field would also be correspondingly small, minimizing artifacts due to resonant Raman effects. This assertion is validated in our work, where we find that the scattering intensities of the $E_g$ modes and the $A_{1g}^{(3)}$ mode are unchanged as a function of out-of-plane magnetic field within the experimental uncertainty, as outlined in Section S3, allowing us to identify magnetophononic zone-folding in the $A_{1g}^{(1)}$ and $A_{1g}^{(2)}$ peaks. Importantly, the phenomena observed in Figs. 1 and 2 are correlated not with magnetic order itself, but specifically with AFM order. The zone-boundary phonon intensity vanishes in the FM phase. In fact, as the results in Fig. 2 show, the zone-boundary intensity of the $A_{1g}^{(2)}$ phonon quantitatively tracks the AFM order in both the temperature- and magnetic field-dependent experiments.

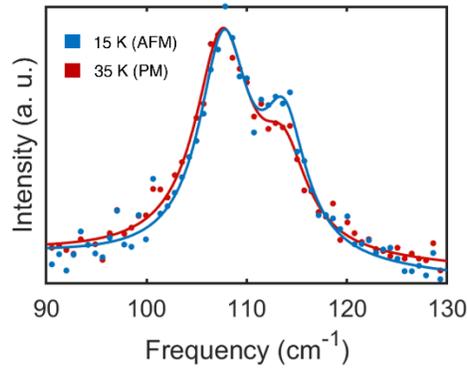

**Fig. S6 Temperature-dependent intensity of $A_{1g}^{(2)}$ mode with 1.58 eV excitation.** Raman spectra measured at 15 K and 35 K using a 1.58 eV (785 nm) laser excitation. The dots are experimental datapoints, and the solid lines are fits as outlined in the Methods section.

Based on the above arguments, in order to rule out resonant Raman effects and isolate peak intensity changes due to magnetophononic coupling, it is essential to measure and analyze phonon scattering intensities as a function of both temperature and magnetic field, as carried out in the present work.

## S6. Symmetry of anomalous scattering intensity

Magnetic ordering can potentially give rise to one-magnon and two-magnon resonances in Raman spectra. In order to eliminate the possibility that the anomalous scattering intensities observed in our work (plotted in Figs. 1 and 2) are due to magnons, we carry out a polarization analysis.

Magnons, by virtue of breaking time-reversal symmetry necessarily have off-diagonal terms in the Raman tensor[5]. In MnBi$_2$Te$_4$, this is associated with $E_g$ modes, as opposed to $A_{1g}$ modes which are fully symmetric and have only diagonal components. The symmetry associated with Raman scattering intensity can be identified as $A_{1g}$ or $E_g$ using polarized Raman measurements, as in Section S1. Here, we focus on the $A_{1g}^{(2)}$ mode. In Fig. S7, we show Raman spectra obtained below and above the AFM ordering temperature $T_N$ = 24 K, corresponding to parallel-polarization, which is sensitive to both $A_{1g}$ and $E_g$ modes, and cross-polarization, which is sensitive only to $E_g$ modes. Our results clearly show that the anomalous scattering intensity overlapped with the $A_{1g}^{(2)}$ phonon in the AFM phase has an $A_{1g}$ symmetry, since it is absent in the cross-polarized channel. This rules out the possibility that it is due to a magnon. It is instead consistent with our interpretation in terms of scattering intensity due to $A_{1g}^{(2)}$ zone-boundary phonons.

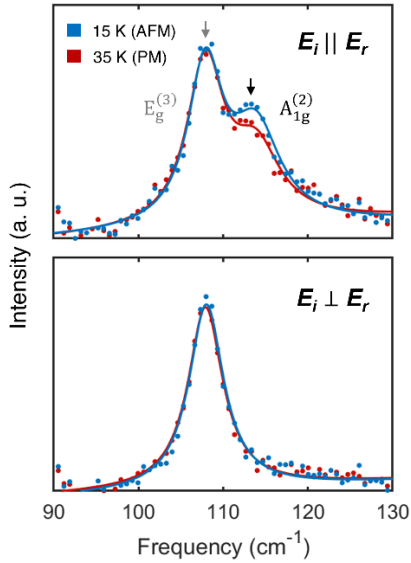

**Fig. S7 Temperature- and polarization-dependent $A_{1g}^{(2)}$ scattering intensity.** Raman spectra measured at 15 K and 35 K with parallel- (top) and cross-polarization (bottom) of incident and reflected light. The dots are experimental datapoints, and solid lines are fits as outlined in the Methods section.

The above inference is also consistent with magnon dispersions measured using inelastic neutron scattering[4]. The dispersion shows that zone-center magnons are at around 1 meV (~8 cm$^{-1}$), whereas the highest energy zone-boundary magnons are at 3 meV (~25 cm$^{-1}$). The low energy of zone-center magnons rules out the possibility of one-magnon resonance interfering with phonon peaks. A two-magnon resonance may plausibly interfere with the $A_{1g}^{(1)}$ mode at 47 cm$^{-1}$ but would be too low in energy to affect the $A_{1g}^{(2)}$ mode at 115 cm$^{-1}$, ruling it out as an explanation for the observed phenomena. Two-magnons are also typically associated with a broad continuum of excitations rather than a well-defined peak, a feature that we do not observe in our experiments.

## S7. Pump-probe measurements – Fluence-dependence

The pump-probe measurements outlined in the main text show phonon excitation via a displacive mechanism, where the ultrafast excitation of carriers by the pump pulse shifts the quasiequilibrium ionic coordinates, generating coherent phonons. Here, the amplitude of coherent phonons is directly proportional to the pump-induced carrier density, i. e. the absorbed fluence. In this context, field-dependent optical conductivity changes may influence coherent phonon amplitudes, in addition to the magnetophononic coupling highlighted in the main text. We account for such magnetic-field dependent changes in absorbed fluence by normalizing the pump-probe measurements with respect to the pump-induced carrier density. The carrier density can be tracked by the maximum amplitude of the transient reflectivity trace, which occurs at a time delay of ~0.9 ps. In Fig. 4 of the main text, all the pump-probe traces are normalized with respect to this amplitude.

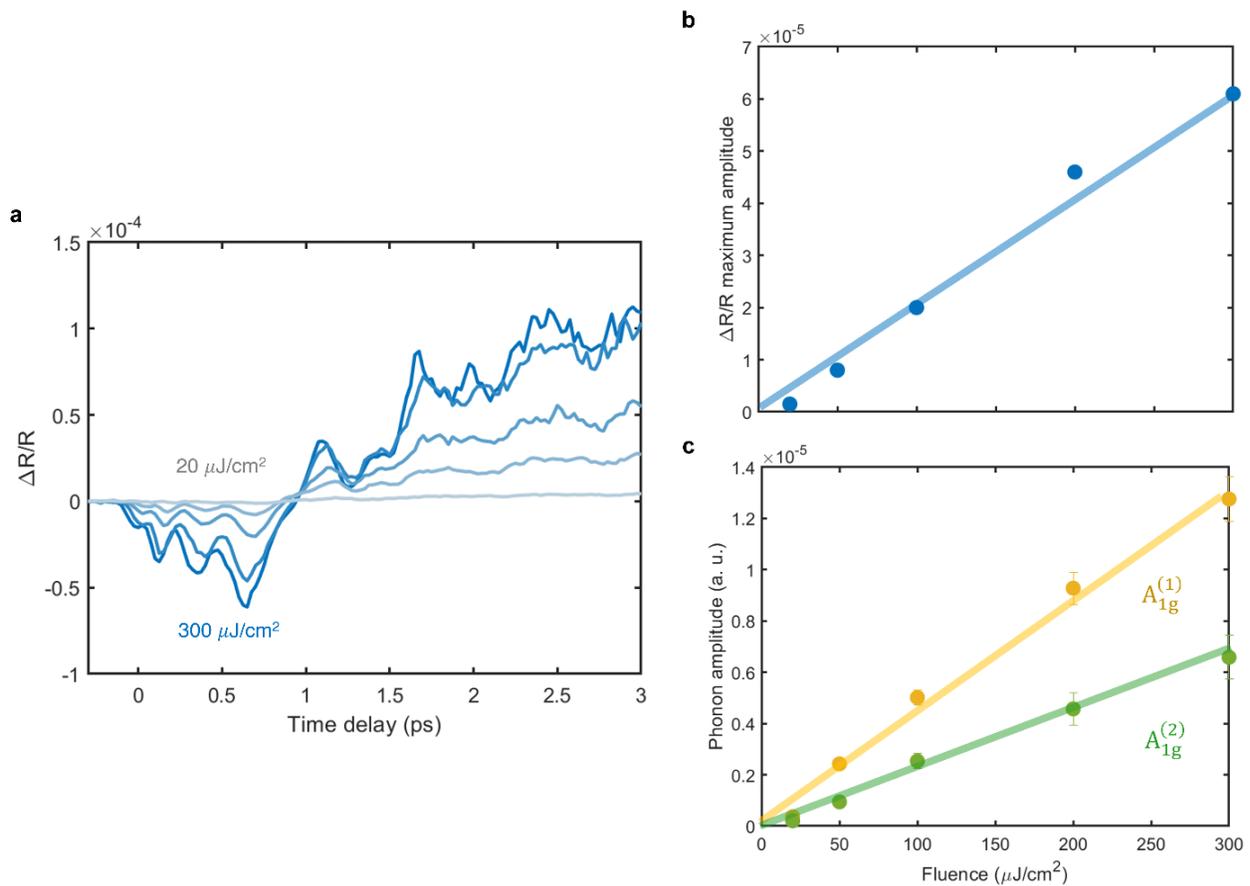

**Fig. S8 Fluence-dependence of coherent phonons. a,** Transient reflectivity traces as a function of pump fluence from 20 to 300 $\mu J/cm^2$. **b,** Maximum sub-picosecond amplitude of transient reflectivity as a function of pump fluence, which is a measure of the photoinduced carrier density. **c,** Amplitude of $A_{1g}^{(1)}$ and $A_{1g}^{(2)}$ coherent phonons as a function of pump fluence, extracted using the method outlined in the main text and Methods. The lines in b and c are linear fits.

We verify the validity of this approach by separately measuring the transient reflectivity and coherent phonon amplitudes as a function of pump fluence, shown in Fig. S8a. The fits in Fig. S8b and S8c, carried out as outlined in the Methods, show that both the maximum transient reflectivity (which tracks the absorbed fluence and photoinduced carrier density) as well as the unnormalized coherent phonon amplitudes scale linearly with fluence, confirming that the maximum amplitude of the transient reflectivity trace indeed tracks the carrier density, validating the normalization procedure used in the main text.

Finally, we note that outside of magnetic-field dependent changes in absorbed fluence, there may potentially be additional magnetodielectric effects that change the electron-phonon interactions and thus the Raman susceptibility, which can affect coherent phonon generation. To lowest order, such changes may be phenomenologically described by a magnetodielectric effect of the form $\chi_e = \chi_e^{(0)} + \gamma M^2$, where $\chi_e$ is the electrical susceptibility, $\gamma$ is magnetodielectric coefficient, and $M$ is the net magnetization. Below, we explore possible changes to electron-phonon interaction due to such a magnetodielectric effect. In the interest of conceptual clarity, we consider a simple $M$-$H$ dependence $M = \chi_m H$, where $\chi_m$ is the magnetic susceptibility and $H$ is the external magnetic field, the Raman susceptibility of a phonon (which determines the coherent phonon amplitude via a Raman-like displacive excitation) can then be written as $\frac{d\chi_e}{du} = \frac{d\chi_e^{(0)}}{du} + H^2 \left(2\gamma\chi_m \frac{d\chi_m}{du} + \chi_m^2 \frac{d\gamma}{du}\right)$. The expression in parenthesis determines the change in Raman susceptibility due to the magnetic field, where $\frac{d\chi_m}{du}$ and $\frac{d\gamma}{du}$ are the phonon modulation of the magnetic susceptibility and magnetodielectric coupling coefficient, respectively. Such a field-dependent change in the coherent phonon amplitude would then be a form of indirect magnetophononic coupling. Based on our current pump-probe experimental data, we cannot completely rule out that such indirect magnetophononic effects also have a contribution, in addition to the direct magnetophononic effects highlighted in our manuscript.

## S8. Debye-Waller effect in ultrafast electron diffraction

Optical pump-probe experiments are typically initiated by ultrafast optical (pump) pulses which generate photo-excited carriers (electrons and holes) – which, after thermalizing, decay through the generation of lattice (and spin) excitations. The lattice excitations result in a disordering of the lattice. Ultrafast electron diffraction is a direct probe of this pump-induced lattice disorder, through the transient Debye-Waller effect.

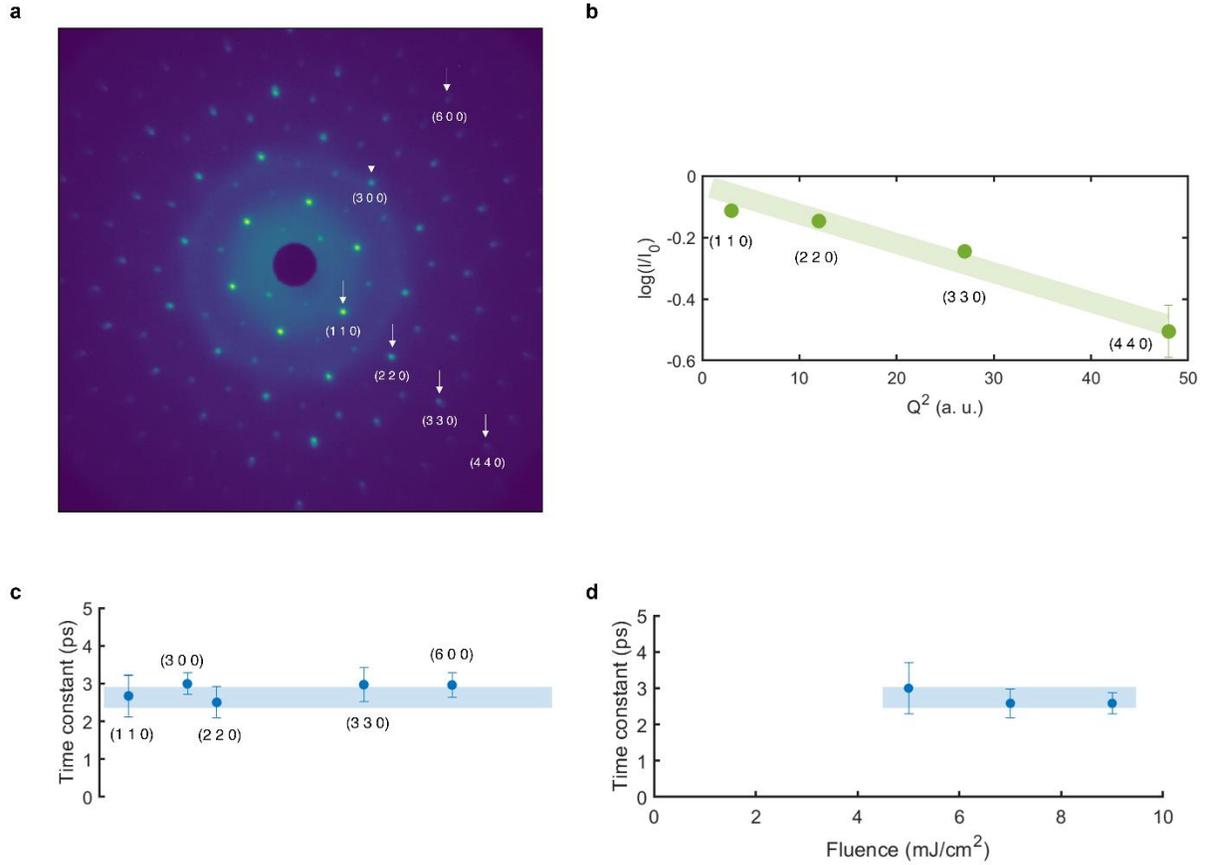

**Fig. S9 Transient Debye-Waller effect. a,** Static electron diffraction image, with the (n n 0) family of peaks labeled. **b,** Transient Bragg reflection intensity obtained in the $t \to \infty$ limit from the exponential decay fit, plotted as a function of $Q$. The green line is a guide to the eye. **c,** The exponential decay constant of the transient intensity of various Bragg peaks. **d,** The exponential decay constant of the (2 2 0) peak as a function of pump fluence. Error bars are standard deviations of fit values. The blue lines are guides to the eye.

In general, the intensity of a Bragg reflection in an electron diffraction experiment is given by

$$I_0(Q) = \sum_j f_j(Q) \exp(-2\pi^2 B_j Q^2) \exp(-i2\pi Q \cdot r_j),$$

where the summation is over atoms in the unit cell (indexed by $j$), $Q$ is the scattering wavevector, $f_j$ is the atomic structure factor for electron diffraction for atom $j$, $B_j$ is the isotropic Debye-Waller factor for atom $j$, and $r_j$ is the position of atom $j$ in the unit cell. The Debye-Waller factor is given by $B_j = \langle u_j^2 \rangle$, which is the root-mean-square displacement of atom $j$ about its mean position. A representative static diffraction pattern from a ~100 nm flake of MnBi$_2$Te$_4$ oriented along the (0 0 1) crystallographic direction is shown in Fig. S9a.

Pump-induced lattice disorder increases $B_j$, i. e. $B_j \to B_j + \Delta B_j$, thus generally resulting in a decrease in the transient Bragg reflection intensities. We can define an effective transient Debye-Waller factor $\Delta B_{eff}$ as follows –

$$I(Q) = \sum_j f_j(Q) \exp(-2\pi^2(B_j + \Delta B_j)Q^2) \exp(-i2\pi Q.r_j)$$

$$= \exp(-2\pi^2 \Delta B_{eff} Q^2) I_0(Q)$$

The transient Bragg reflection intensity $I(Q)$ is thus a quantitative measure of the pump-induced lattice disorder, $\Delta B_{eff} = \Delta \langle u_{eff}^2 \rangle$. The Debye-Waller effect for a given family of Bragg reflections scales with $Q$, such that $\log\left(\frac{I}{I_0}\right) \propto -Q^2$.

In Fig. S9b, we plot the transient intensity $\log\left(\frac{I}{I_0}\right)$ of the (*n n 0*) family of Bragg reflections at the $t \to \infty$ limit from the exponential decay fit (see Methods), as a function of $Q^2$. The linear dependence confirms that the observed evolution of transient Bragg intensities is due to a Debye-Waller effect via pump-induced lattice disorder.

The time constants of the transient intensities of various Bragg peaks, obtained from exponential decay fits, are shown in Fig. S9c. Within the experimental uncertainty, the time constant is uniform across different peaks.

In the main text, we use the transient Debye-Waller time constant from our UED measurements to establish the timescale of lattice thermalization. However, the optical pump-probe measurements reported in the main text use a much lower fluence, of 0.1 mJ/cm$^2$, as opposed to 7 mJ/cm$^2$ used in the UED measurements reported above. In this context, we report the thermalization time constants from our UED measurements as a function of fluence, in Fig. S9d. The time constants are largely unchanged from 5 to 9 mJ/cm$^2$, with a slight increase at lower fluences.

Such a behavior is consistent with increased phonon-phonon scattering at higher fluences[6]. It is expected then that the thermalization time constant at the low fluences used in our optical measurements, with their correspondingly lower phonon populations, would likely be even higher than that extracted from the UED measurements; i. e. the UED time constant sets a lower bound for the phonon thermalization time. This supports our assertion that phonon subsystem remains in a nonequilibrium state through the entire time delay range measured in our study.